\documentstyle[12pt]{article}
\newcommand{\ltwid}{\raise.3ex\hbox{$<$\kern-.75em\lower1ex\hbox{$\sim
$}}}

\title{{\hfill\normalsize ITP-95-02E}\\[1.0cm]
Dynamical Chiral Symmetry Breaking by a Magnetic Field in QED}

\author{{\sl V.P. Gusynin , V.A. Miransky  and I.A. Shovkovy}\\
{\sl Bogolyubov Institute for Theoretical Physics,
252143 Kiev, Ukraine}}
\date{}
\begin{document}
\maketitle

\vfill

\begin{abstract}
It is shown that the chiral symmetry is spontaneously broken by a constant
magnetic field in QED. The dynamical mass of fermions (energy gap in the
fermion spectrum) is $m_{dyn}\simeq
C\sqrt{eB}\exp\left[-\left(\pi/\alpha\right)
^{1/2}\right]$, where $B$ is the magnetic field, the constant $C$ is of
order one and $\alpha=e^2/4\pi$ is the renormalized coupling constant.
Possible applications of this effect are discussed.
\end{abstract}

\vfill
\eject

The dynamics of fermions in an external constant magnetic field in QED
was considered by Schwinger long ago \cite{Sch}. In those classical works,
while the interaction with the external field was considered in all orders
in the coupling constant, the quantum dynamics was treated perturbatively.
There is no dynamical chiral symmetry breaking in QED in this approximation
\cite{CalChod}. In this paper we reconsider
this problem, treating the QED dynamics nonperturbatively. We will show
that a constant magnetic field $B$ leads to dynamical chiral symmetry
breaking in massless QED. The dynamical mass of fermions (energy gap in
the fermion spectrum) is:
\begin{equation}
m_{dyn}\simeq C\sqrt{eB}\exp{\left[-\left(\frac{\pi}{\alpha}\right)^{1/2}
\right]},\label{eq:mdyn}
\end{equation}
where the constant $C$ is of order one and $\alpha=e^2 /4\pi$ is the
renormalized coupling constant relating to the scale $\mu=m_{dyn}$.

The essence of this effect is the dimensional reduction $D\to D-2$ ($3+1\to
1+1$ in this case) in the dynamics of fermion pairing in a magnetic field,
pointed out recently in Refs.\cite{GMS1,GMS2}. Actually, we will show that,
in Euclidean space, the equation describing the Nambu--Goldstone (NG) bosons
in QED in a magnetic field has the form of a two--dimensional Schr\"{o}dinger
equation:
\begin{equation}
(-\Delta+m^2_{dyn}+V({\bf r}))\Psi({\bf r})=0.\label{eq:schre}
\end{equation}
Here $\Psi({\bf r})$ is expressed through the Bethe--Salpeter (BS) function
of NG bosons, $\Delta=\partial^2/\partial x^2_3+\partial^2/\partial x^2_4$
(the magnetic field is in the $+x_3$ direction; $x_4=it$), and
the potential $V({\bf r})$ is
\begin{equation}
V({\bf r})=\frac{\alpha}{\pi l^2}\exp{\left(\frac{r^2}{2l^2}\right)}
Ei\left(-\frac{r^2}{2l^2}\right),\qquad r^2=x^2_3+x^2_4,
\end{equation}
where $Ei(x)=-\int_{-x}^{\infty}dt \exp(-t)/t$ is the integral exponential
function \cite{Ryz} and $l\equiv |eB|^{-1/2}$ is the magnetic length.

We emphasize that we work in the conventional, weak coupling, phase of
QED. That is, the bare coupling $\alpha^{(0)}$, relating to the scale
$\mu=\Lambda$, where $\Lambda$ is an ultraviolet cutoff, is assumed to be
small, $\alpha^{(0)}\ll 1$. Then, because of infrared freedom in QED,
interactions in the theory are weak at all scales and, as a result, the
treatment of the nonperturbative dynamics is reliable.

We will consider possible applications of this effect at the end of the paper.

The Lagrangian density of massless QED in a magnetic field is:
\begin{equation}
{\cal L} =-\frac{1}{4}F^{\mu\nu}F_{\mu\nu}+
\frac{1}{2} \left[\bar{\psi},(i\gamma^\mu D_\mu)\psi\right],
\label{eq:lag}
\end{equation}
where the covariant derivative $D_\mu$ is
\begin{equation}
D_\mu=\partial_\mu-ie(A^{ext}_\mu+A_\mu),\qquad A^{ext}_\mu =
\left(0,-\frac{B}{2} x_2 ,\frac{B}{2} x_1,0\right).\label{eq:vecA}
\end{equation}
Besides the Dirac index $(n)$, the fermion field carries an additional,
flavor, index $a=1,2,\dots,N$. Then, at $N\geq 2$, the Lagrangian
density (\ref{eq:lag}) is invariant under the chiral
$SU_L(N)\times SU_R(N)\times U_{L+R}(1)$ transformations
( we will not discuss the dynamics related to the anomalous, singlet,
current $j_{5\mu}$ in this paper).
Since we consider the weak coupling phase of QED, there is no
spontaneous chiral symmetry breaking at $B=0$ \cite{Fomin}. We will show
that the magnetic field changes the situation dramatically: at $B\neq 0$
the chiral symmetry $SU_L(N)\times SU_R(N)$ breaks down to the
$SU_V(N)\equiv SU_{R+L}(N)$ and, as a result, there appear $N^2-1$ gapless NG
bosons composed of fermions and antifermions. As we will see, the BS equation
for the NG bosons defines the dynamical mass (energy gap) for fermions.

The homogeneous BS equation for $N^2-1$ NG bound states takes
the form (for a review see Ref.\cite{Mir}):
\begin{eqnarray}
\chi_{AB}^\beta(x,y;P)&=&-i\int d^4x_1 d^4y_1 d^4x_2 d^4y_2
G_{AA_{1}}(x,x_1)
K_{A_{1}B_{1};A_{2}B_{2}}(x_1y_1,x_2y_2)\nonumber\\
&\cdot&\chi_{A_{2}B_{2}}^\beta(x_2,y_2;P)G_{B_1B}(y_1,y),
\end{eqnarray}
where the BS wave function $\chi^{\beta}_{AB}(x,y;P)=\langle 0|T\psi_A(x)
{\bar\psi}_B(y)|P;\beta \rangle$, $\beta=1,2,\dots,N^2-1$, and
the fermion propagator $G_{AB}(x,y)=
\langle 0|T\psi_A(x){\bar\psi}_B(y)|0 \rangle$; the indices $A=(na)$
and $B=(mb)$ include both Dirac $(n,m)$ and flavor $(a,b)$
indices. Note that though the external field $A_\mu^{ext}$ (\ref{eq:vecA})
breaks the
conventional translation invariance, the total momentum $P$ is a good,
conserved, quantum number for neutral channels \cite{Avron}, in particular,
for these NG bosons. Since, as will be shown below, the NG bosons are
formed in the infrared region, where the QED coupling  is weak, one
can use the BS kernel in leading order in $\alpha$  \cite{Mir}:
\begin{eqnarray}
K_{A_{1}B_{1};A_{2}B_{2}}(x_1y_1,x_2y_2)&=&-4\pi i\alpha\delta_{a_1a_2}
\delta_{b_2b_1}\gamma^\mu_{n_1n_2}\gamma^\nu_{m_2m_1}{\cal D}_{\mu\nu}
(y_2-x_2)\delta(x_1-x_2)\delta(y_1-y_2) \nonumber \\
&+&4\pi i\alpha\delta_{a_1b_1}\delta_{b_2a_2}
\gamma^\mu_{n_1m_1}\gamma^\nu_{m_2n_2}
{\cal D}_{\mu\nu}(x_1-x_2)\delta(x_1-y_1)\delta(x_2-y_2),\label{eq:ker}
\end{eqnarray}
where the photon propagator
\begin{equation}
{\cal D}_{\mu\nu}(x)=\frac{-i}{(2\pi)^4}\int d^4k e^{ikx} \left(g_{\mu\nu}-
\lambda\frac{k_\mu k_\nu}{k^2}\right)\frac{1}{k^2}\label{eq:dmunu}
\end{equation}
($\lambda$ is a gauge parameter). The first term on the right--hand
side of Eq.(\ref{eq:ker}) corresponds to the ladder
approximation. The second (annihilation) term does not contribute to the BS
equation for massless NG bosons (this follows from the fact that, due to
the Ward identities for axial currents, the BS equation for the NG bosons can
be reduced to the Schwinger--Dyson equation for the fermion propagator where
there is no contribution of the annihilation term \cite{Mir}). Therefore we
shall omit this term in the following. Then the BS equation takes the form:
\begin{eqnarray}
\chi^\beta_{AB}(x,y;P)&=&-4\pi\alpha\int d^4x_1 d^4y_1 S_{AA_1}(x,x_1)
\delta_{a_1a_2}\gamma^\mu_{n_1n_2}\chi^\beta_{A_2B_2}(x_1,y_1;P) \nonumber \\
&\cdot&\delta_{b_2b_1}\gamma^\nu_{m_2m_1}S_{B_1B}(y_1,y)
{\cal D}_{\mu\nu}(y_1-x_1),  \label{eq:bs1}
\end{eqnarray}
where, since we are working to lowest order in $\alpha$, the full
fermion propagator $G_{AB}(x,y)$ is replaced by the propagator $S$ of a free
fermion (with the mass $m=m_{dyn}$) in a magnetic field \cite{Sch}:
\begin{equation}
S_{AB}(x,y) =\delta_{ab} \exp \left(\frac{ie}{2}(x-y)^\mu A_\mu^{ext}
(x+y)\right)\tilde{S}_{nm} (x-y),
\end{equation}
where the Fourier transform of $\tilde{S}$ is
\begin{eqnarray}
\tilde{S}(k) &=&\int\limits^\infty_0 ds \exp \left[is\left(k^2_0-k^2_3
-{\bf k}^2_{\perp}\frac{\tan(eBs)}{eBs}-m_{dyn}^2\right)\right] \nonumber \\
&\cdot & \left[(k^0\gamma^0-k^3\gamma^3+m_{dyn})(1+\gamma^1\gamma^2\tan(eBs))
-{\bf k}_{\perp}\mbox{\boldmath$\gamma$}_{\perp}(1+\tan^2(eBs))\right]
\end{eqnarray}
(here ${\bf k}_{\perp}=(k_1,k_2)$, $\mbox{\boldmath$\gamma$}_{\perp}=
(\gamma_1,\gamma_2)$). Using the new variables, the center mass coordinate,
$R=(x+y)/2$, and the relative coordinate, $r=x-y$, equation (\ref{eq:bs1})
can be rewritten as
\begin{eqnarray}
\tilde{\chi}_{nm}(R,r;P)&=&-4\pi\alpha\int d^4R_1 d^4r_1 \tilde{S}_{nn_1}
\left(R-R_1+\frac{r-r_1}{2}\right)\gamma^\mu_{n_1n_2}
\tilde{\chi}_{n_2m_2}(R_1,r_1;P) \gamma^\nu_{m_2m_1} \nonumber\\
&\cdot&\!\!\!\tilde{S}_{m_1m}\left(\frac{r-r_1}{2}-R+R_1\right)\!
{\cal D}_{\mu\nu}(-r_1)\exp\left[-ie(r+r_1)^\mu A_\mu^{ext}
(R-R_1)\right]\!. \label{eq:bs12}
\end{eqnarray}
Here the function $\tilde{\chi}_{nm}(R,r;P)$ is defined from the equation
\begin{equation}
\chi^{\beta}_{AB}(x,y;P)=\langle 0|T\psi_A(x){\bar\psi}_B(y)|P;\beta\rangle
=\lambda^\beta_{ab} \exp[ier^\mu A_\mu^{ext}(R)]\tilde{\chi}_{nm}(R,r;P)
\label{eq:bsfun}
\end{equation}
where $\lambda^\beta$ are $N^2-1$ flavor matrices ($tr(\lambda^\beta
\lambda^\gamma)=2\delta_{\beta\gamma}$; $\beta$,$\gamma=1,2,\dots,N^2-1$).
The important fact is that the effect of translation symmetry breaking by the
external field is factorized in the phase factor in Eq.(\ref{eq:bsfun})
and equation (\ref{eq:bs12}) admits a translation invariant solution,
$\tilde{\chi}_{nm}(R,r;P)=\exp(-iPR)\tilde{\chi}_{nm}(r;P)$.
Henceforth we will consider the case with $P\to 0$.
Then, transforming this equation into momentum space, we get:
\begin{eqnarray}
\tilde{\chi}_{nm}(p)&=&-4\pi\alpha\int \frac{d^2q_{\perp} d^2R_{\perp}
d^2k_{\perp} d^2k_{\parallel}}{(2\pi)^6} \exp(-i{\bf q}_{\perp}{\bf R})
\tilde{S}_{nn_1}\left(p_{\parallel},{\bf p}_{\perp}+e{\bf A}^{ext}(R)+
\frac{{\bf q}_{\perp}}{2}\right) \nonumber\\
&\cdot&\gamma^\mu_{n_1n_2}\tilde{\chi}_{n_2m_2}(k)
\gamma^\nu_{m_2m_1}\tilde{S}_{m_1m}\left(p_{\parallel},{\bf p}_{\perp}+
e{\bf A}^{ext}(R)-\frac{{\bf q}_{\perp}}{2}\right) \nonumber\\
&\cdot&{\cal D}_{\mu\nu}(k_{\parallel}-p_{\parallel},{\bf k}_{\perp}-
{\bf p}_{\perp}-2e{\bf A}^{ext}(R)),\label{eq:bs2}
\end{eqnarray}
where $p_{\parallel}\equiv (p^0,p^3)$, ${\bf p}_{\perp}\equiv (p^1,p^2)$.

The crucial point for the further analysis will be the assumption that
$m_{dyn}\ll \sqrt{|eB|}$ and that the region mostly responsible for
generating the mass is the infrared region with $k\ltwid m_{dyn}\ll
\sqrt{|eB|}$. As we shall see below, this assumption is self--consistent
(see Eq.(\ref{eq:mdyn})). The assumption allows us to replace the propagator
$\tilde{S}_{nm}$ in Eq.(\ref{eq:bs2}) by the pole contribution of the lowest
Landau
level (LLL). In order to show this, we recall that the energy spectrum of
fermions with $m=m_{dyn}$ in a magnetic field is \cite{Akh}
\begin{equation}
E_n(p_3) = \pm\sqrt{m^2_{dyn}+2|eB|n+p^2_3},\ \qquad  n=0,1,2,\dots .
\end{equation}
(the Landau levels). The  propagator $\tilde{S}(p)$ can be decomposed
over the Landau level poles \cite{GMS2,Cho}:
\begin{equation}
\tilde{S}(p)=i \exp\left(-\frac{{\bf p}^2_{\perp}}{|eB|}\right)
\sum_{n=0}^{\infty}(-1)^n\frac{D_n(eB,p)}{p_0^2-p_3^2-m_{dyn}^2-2|eB|n}
\label{eq:poles}
\end{equation}
with
\begin{eqnarray*}
D_n(eB,p)&=&(p^0\gamma^0-p^3\gamma^3+m_{dyn})\Bigg[(1-i\gamma^1\gamma^2)
L_n\left(2\frac{{\bf p}^2_{\perp}}{|eB|}\right) \nonumber \\
&-&(1+i\gamma^1\gamma^2)L_{n-1}\left(2\frac{{\bf p}^2_{\perp}}{|eB|}\right)
\Bigg]+4(p^1\gamma^1+p^2\gamma^2)L_{n-1}^1
\left(2\frac{{\bf p}^2_{\perp}}{|eB|}\right),
\end{eqnarray*}
where $L_n(x)$ are the generalized Laguerre polynomials ($L_n\equiv L_n^0$
and $L_{-1}^{\alpha}(x)=0$ by definition).  Eq.(\ref{eq:poles})
implies that at $m_{dyn}\ll \sqrt{|eB|}$, the LLL with $n=0$ dominates in
the infrared region with $p\ltwid m_{dyn}$.

The LLL pole dominance essentialy simplifies the analysis. Now
\begin{eqnarray}
\tilde{S}(p)&\simeq &i\exp(-l^2{\bf p}^2_{\perp})\frac{\hat{p}_{\parallel}+
m_{dyn}}{p^2_{\parallel}-m^2_{dyn}}(1-i\gamma^1\gamma^2),\label{eq:0lev}
\end{eqnarray}
where $\hat{p}_{\parallel}=p^0\gamma^0-p^3\gamma^3$ and
$\hat{p}_{\parallel}^2=(p^0)^2-(p^3)^2$, and equation (\ref{eq:bs2})
transforms into the following one:
\begin{eqnarray}
\rho(p_{\parallel},{\bf p}_{\perp})&=&\frac{2\alpha l^2}{(2\pi)^4}
e^{-l^2{\bf p}^2_{\perp}}\int d^2A_{\perp} d^2k_{\perp}d^2k_{\parallel}
e^{-l^2{\bf A}^2_{\perp}}(1-i\gamma^1\gamma^2)\gamma^\mu
\frac{\hat{k}_{\parallel}+m_{dyn}}{k^2_{\parallel}-m^2_{dyn}} \nonumber\\
&\cdot&\rho(k_{\parallel},{\bf k}_{\perp})\frac{\hat{k}_{\parallel}+m_{dyn}}
{k^2_{\parallel}-m^2_{dyn}}\gamma^\nu (1-i\gamma^1\gamma^2)
{\cal D}_{\mu\nu}(k_{\parallel}-p_{\parallel},{\bf k}_{\perp}-
{\bf A}_{\perp}),\label{eq:rho}
\end{eqnarray}
where
\begin{equation}
\rho(p_{\parallel},{\bf p}_{\perp})=(\hat{p}_{\parallel}-m_{dyn})
\tilde{\chi}(p)(\hat{p}_{\parallel}-m_{dyn}).
\end{equation}
Eq.(\ref{eq:rho}) implies that $\rho(p_{\parallel},{\bf p}_{\perp})=
\exp(-l^2{\bf p}^2_{\perp})\varphi(p_{\parallel})$, where
$\varphi(p_{\parallel})$ satisfies the equation:
\begin{eqnarray}
\varphi(p_{\parallel})&=&\frac{\pi\alpha}{(2\pi)^4}\int d^2k_{\parallel}
(1-i\gamma^1\gamma^2)\gamma^\mu\frac{\hat{k}_{\parallel}+m_{dyn}}
{k^2_{\parallel}-m^2_{dyn}}\varphi(k_{\parallel}) \nonumber\\
&\cdot&\frac{\hat{k}_{\parallel}+m_{dyn}}
{k^2_{\parallel}-m^2_{dyn}}\gamma^\nu (1-i\gamma^1\gamma^2)
{\cal D}^{\parallel}_{\mu\nu}(k_{\parallel}-p_{\parallel}), \label{eq:phi}
\end{eqnarray}
\begin{equation}
{\cal D}^{\parallel}_{\mu\nu}(k_{\parallel}-p_{\parallel})=
\int d^2k_{\perp}\exp(-\frac{l^2{\bf k}^2_{\perp}}{2})
{\cal D}_{\mu\nu}(k_{\parallel}-p_{\parallel},{\bf k}_{\perp}).
\end{equation}
Thus the BS equation has been reduced to a two--dimensional integral
equation. Of course, this fact reflects the two--dimensional character of
the dynamics of the LLL in the infrared region, that can be explicitly
read from Eq.(\ref{eq:0lev}).

We emphasize that the dimensional reduction  in a magnetic field does not
affect the dynamics of the center of mass of {\em neutral} bound
states (in particular, these $(N^2-1)$ NG bosons). Indeed, the reduction
$3+1\to 1+1$ in the fermion propagator, in the infrared region, reflects
the fact that the motion of {\em charged} particles is restricted in
directions perpendicular to the magnetic field. Since there is no such
a restriction for the motion of the center of mass of neutral particles,
their propagator must have a $(3+1)$--dimensional form \cite{Note}. This
fact was explicitly shown for neutral bound states in the
Nambu--Jona--Lasinio model in a magnetic field, in $1/N_c$ expansion
\cite{GMS2} and for neutral excitations in nonrelativistic systems
\cite{Cond}. Since, besides that, the propagator of {\em massive}
fermions is nonsingular at small momenta, we conclude
that the infrared dynamics of the NG modes is soft in the present model.
This in particular implies that the phenomenon of spontaneous
chiral symmetry breaking in this model does not contradict to the
Mermin--Wagner--Coleman theorem \cite{Mermin} forbidding the spontaneous
breakdown of continuous symmetries at $D=1+1$.

Henceforth we will use Euclidean space with $k_4=-ik^0$, where the total
momentum $P$ of NG bosons equals zero. In order to define the matrix
structure of the wave function $\varphi(p_{\parallel})$ of the NG bosons,
note that, in a magnetic field, there is the symmetry $SO(2)\times SO(2)
\times {\cal P}$, where the $SO(2)\times SO(2)$ is connected with rotations
in $x_1$--$x_2$ and $x_3$--$x_4$ planes and ${\cal P}$ is the inversion
transformation
$x_3\to -x_3$ (under which a fermion field transforms as $\psi\to i\gamma_5
\gamma_3\psi$). This symmetry implies that the function
$\varphi(p_{\parallel})$ takes the form:
\begin{equation}
\varphi(p_{\parallel})=\gamma_5 (A+i\gamma_1\gamma_2B+\hat{p}_{\parallel}C+
i\gamma_1\gamma_2\hat{p}_{\parallel}D)   \label{eq:phii}
\end{equation}
where $\hat{p}_{\parallel}=p_3\gamma_3+p_4\gamma_4$ and $A,B,C$ and $D$ are
functions of $p^2_{\parallel}$ ($\gamma_\mu$ are antihermitian in Euclidean
space).

We begin the analysis of the equation (\ref{eq:phi}) by choosing the
Feynman gauge (the general covariant gauge will be considered below). Then,
\begin{eqnarray*}
{\cal D}^{\parallel}_{\mu\nu}(k_{\parallel}-p_{\parallel})=i\delta_{\mu\nu}\pi
\int\limits^{\infty}_{0}\frac{dx \exp(-l^2x/2)}{(k_{\parallel}-
p_{\parallel})^2+x}
\end{eqnarray*}
and, substituting expansion (\ref{eq:phii}) into equation (\ref{eq:phi}),
we find that $B=-A$, $C=D=0$, {\em i.e.},
$\varphi(p_{\parallel})=A\gamma_5(1-i\gamma_1\gamma_2)$,
and the function $A$ satisfies the equation
\begin{equation}
A(p)=\frac{\alpha}{2\pi^2}\int\frac{d^2k A(k)}{k^2+m^2_{dyn}}
\int\limits^{\infty}_{0} \frac{dx \exp(-xl^2/2)}{({\bf k-p})^2+x}
\label{eq:Ap}
\end{equation}
(henceforth we will omit the symbol $\parallel$ from $p$ and $k$).
Introducing the function
\begin{eqnarray*}
\Psi({\bf r})=\int\frac{d^2k}{(2\pi)^2}\frac{A(k)}{k^2+m_{dyn}^2}
e^{i{\bf kr}},
\end{eqnarray*}
we get (from Eq.(\ref{eq:Ap})) the two--dimensional Schr\"{o}dinger
equation (\ref{eq:schre}) with the potential
\begin{eqnarray}
V({\bf r})&=&-\frac{\alpha}{2\pi^2}\int d^2p e^{i{\bf pr}}
\int\limits^{\infty}_{0} \frac{dx \exp(-x/2)}{l^2p^2+x}=
-\frac{\alpha}{\pi l^2}\int\limits^{\infty}_{0} dx e^{-x/2}
K_{0}\left(\frac{r}{l}\sqrt{x}\right) \nonumber\\
&=&\frac{\alpha}{\pi l^2} \exp\left(\frac{r^2}{2l^2}\right)
Ei\left(-\frac{r^2}{2l^2}\right)\label{eq:pot}
\end{eqnarray}
($K_0$ is the Bessel function). Since $-m_{dyn}^2$ plays the
role of energy $E$ in this equation and $V({\bf r})$ is a negative,
{\em i.e.} attractive, potential, the problem is reduced to finding the
spectrum of bound states (with $E=-m_{dyn}^2<0$) of the two--dimensional
Schr\"{o}dinger equation with the potential (\ref{eq:pot}). For this
purpose, we can use
some results proved in the literature. First, the energy of the lowest level
$E(\alpha)$ for the two--dimensional Schr\"{o}dinger equation is a
nonanalytic function of the coupling constant $\alpha$ at $\alpha=0$
\cite{Simon}. Second, if the potential $V({\bf r})$ were short--range, then
$m^2_{dyn}(\alpha)=-E(\alpha)$ would take the form $m^2_{dyn}\sim
\exp[-1/(a\alpha)]$ where $a$ is a positive constant \cite{Simon}.
However, the potential is long--range in our case. Indeed, using the
asymptotic relations for $Ei(x)$ \cite{Ryz}, we get:
\begin{eqnarray}
V({\bf r})&\simeq &-\frac{2\alpha}{\pi}\frac{1}{r^2}, \qquad r\to\infty;
\nonumber\\
V({\bf r})&\simeq &-\frac{\alpha}{\pi l^2}\left(\gamma+
\ln\frac{2l^2}{r^2}\right), \qquad r\to 0
\end{eqnarray}
where $\gamma\simeq 0.577$ is the Euler constant. To find
$m^2_{dyn}(\alpha)$, we shall use the integral equation (\ref{eq:Ap})
at $p=0$. Then, as $\alpha\to 0$, the dominant contribution in the
integral on the right hand side of Eq.(\ref{eq:Ap}) is formed in the
infrared region with soft $k^2\ltwid m^2_{dyn}$:
\begin{equation}
A(0)\simeq \frac{\alpha}{2\pi^2}A(0)\int\frac{d^2k}{k^2+m^2_{dyn}}
\int\limits^{\infty}_{0} \frac{dx \exp(-y/2)}{l^2k^2+y}\sim
\frac{\alpha}{4\pi}A(0)\left[\ln\left(\frac{m_{dyn}^2l^2}{2}\right)\right]^2,
\end{equation}
{\em i.e.},
\begin{equation}
m_{dyn}=C\sqrt{|eB|}\exp\left[-\sqrt{\frac{\pi}{\alpha}}\right],
\label{eq:m-dyn}
\end{equation}
where the constant $C=O(\alpha^0)$ remains undefined in this approximation.
Note that this result agrees with the analysis of Ref.\cite{Perelomov} where
the analytic properties of $E(\alpha)$ were considered for the Schr\"{o}dinger
equations with potentials having the asymptotics $V(r)\to 1/r^2$ as
$r\to \infty$.

Let us now turn to considering the general covariant gauge (\ref{eq:dmunu}).
As is known,
the ladder approximation is not gauge invariant. However, let us show that,
because the present effect is due to the infrared dynamics in QED, where
the coupling constant is small, the leading term in $\ln(m_{dyn})$,
$\ln(m_{dyn})\simeq -(\pi/\alpha)^{1/2}$, is the same in all covariant gauges.

Acting in the same way as before, we find that in the general covariant
gauge the wave function (\ref{eq:phii}) takes the form
$\varphi(p)=\gamma_5(1-i\gamma_1\gamma_2)(A+\hat{p}C)$ where the functions
$A$ and $C$ satisfy the equations:
\begin{equation}
A(p)=\frac{\alpha}{2\pi^2}\int\frac{d^2k A(k)}{k^2+m^2_{dyn}}
\int\limits^{\infty}_{0} \frac{dx (1-\lambda xl^2/4) \exp(-xl^2/2)}
{({\bf k-p})^2+x}, \label{eq:App}
\end{equation}
\begin{equation}
C(p)=\frac{\alpha\lambda}{4\pi^2}\int\frac{d^2k C(k)}{k^2+m^2_{dyn}}
\left[2k^2-({\bf kp})-\frac{k^2({\bf kp})}{p^2}\right]
\int\limits^{\infty}_{0} \frac{dx \exp(-xl^2/2)}{[({\bf k-p})^2+x]^2}.
\end{equation}
One can see that the dominant contribution on the right-hand side of
Eq.(\ref{eq:App})
(proportional to $[\ln m^2_{dyn}l^2]^2$ and formed at small $k^2$) is
independent of the gauge parameter $\lambda$. Therefore the leading
singularity in $\ln(m_{dyn})$, $\ln(m_{dyn})=-(\pi/\alpha)^{1/2}$, is indeed
gauge invariant.

This concludes the derivation of Eqs.(\ref{eq:mdyn}) and (\ref{eq:schre})
describing spontaneous
chiral symmetry breaking by a magnetic field in QED.

In conclusion, let us discuss possible applications of this effect. One
potential application is the interpretation of the results of the GSI
heavy--ion scattering experiments \cite{Sal} in which narrow peaks are seen
in the energy spectra of emitted $e^{+}e^{-}$ pairs. One proposed
explanation \cite{Celenza} is that a very strong electromagnetic field,
created by the heavy ions, induces a phase transition in QED to a phase with
spontaneous chiral symmetry breaking. The observed peaks are due to the decay
of positronium--like states in this phase. The catalysis of chiral symmetry
breaking by a magnetic field in QED, studied in this paper, can serve as a toy
example of such a phenomenon. In order to get a more realistic model, it would
be interesting to extend this analysis to non--constant background fields
\cite{Caldi}.

Another, potentialy interesting, application can be connected with the
possibility of the existence of very strong magnetic fields
($B\sim 10^{24}$Gauss) during the electroweak phase transition in the
early Universe \cite{Univ}. As the present results suggest, such
fields might essentially change the character of the electroweak phase
transition.

Yet another application of the effect can be connected with the role of
iso-- and chromomagnetic backgrounds as models for the QCD vacuum (the
Copenhagen vacuum \cite{Niel}). Also, as has been suggested recently
\cite{DESY}, isomagnetic fields in the vacuum of electroweak left--right
models can induce the parity breakdown. Our work suggests that such field
configurations may play the important role in triggering chiral symmetry
breaking in QCD and those left--right models.

The work of I.A.Sh. was supported in part by the International Soros Science
Education Program (ISSEP) through grant No.PSU052143.

\begin{center}
Note added in proof:
\end{center}

We have just finished a complete (both analytical and numerical)
analysis of integral equation (\ref{eq:Ap}). The result for $m_{\rm dyn}$
agrees quite well with estimate (\ref{eq:m-dyn}) and is
$m_{\rm dyn}=C\sqrt{|eB|}\exp\left[-\pi/2\sqrt{\pi/2\alpha}\right]$,
where $C$ is of  $O(\alpha^0)$. Notice that the ratio of the powers
of this exponent and that in Eq.(\ref{eq:m-dyn}) is $\pi/2\sqrt{2}\simeq 1.1$.


\begin{thebibliography}{99}

\bibitem{Sch} J. Schwinger, {\sl Phys. Rev.} {\bf 82}, 664 (1951);
{\sl Phys. Rev.} {\bf D7}, 1696 (1973).

\bibitem{CalChod} D.Caldi, A.Chodos, K.Everding, D.Owen, and S.Vafaeisefat,
{\sl Phys. Rev.} {\bf D39}, 1432 (1989).

\bibitem{GMS1} V.P.Gusynin, V.A.Miransky and I.A.Shovkovy,
{\sl Phys. Rev. Lett.} {\bf 73}, 3499 (1994); {\sl Phys. Rev.} {\bf
D52}, 4718 (1995).

\bibitem{GMS2} V.P.Gusynin, V.A.Miransky and I.A.Shovkovy,
{\sl Phys. Lett.} {\bf B349}, 477 (1995).

\bibitem{Ryz} I.S. Gradshtein and I.M. Ryzhik, {\em Table of Integrals,
Series and Products} (Academic Press, Orlando, 1980).

\bibitem{Fomin} P.I.Fomin, V.P.Gusynin, V.A.Miransky, and Yu.A.Sitenko,
{\sl Riv. del Nuov. Cim.} {\bf 6} N5 (1983).

\bibitem{Mir} V.A. Miransky, {\sl Dynamical Symmetry Breaking in Quantum
Field Theories\/} (World Scientific Co., Singapore, 1993).

\bibitem{Avron} J.E.Avron, I.W.Herbst and B.Simon, {\sl Ann. Phys.} {\bf 114},
431 (1978).

\bibitem{Akh} A.I. Akhiezer and V.B. Berestetsky, {\em Quantum
Electrodynamics} (Interscience, NY, 1965).

\bibitem{Cho} A.Chodos, K.Everding and D.A.Owen, {\sl Phys. Rev.} {\bf D42},
2881 (1990).

\bibitem{Note} The Lorentz invariance is broken by a magnetic field in
this problem. By the $(3+1)$--dimensional form, we understand that the
denominator of the propagator depends on energy and all the components
of momentum:
$D(P)\sim (P_0^2-C_{\perp}P_{\perp}^{2}-C_{\parallel}P_{\parallel}^2)^{-1}$
($D(P)\sim (P_0-C_{\perp}P_{\perp}^{2}-C_{\parallel}P_{\parallel}^2)^{-1}$
in nonrelativistic systems) with $C_{\perp},C_{\parallel}\neq 0$.

\bibitem{Cond} B.R. Johnson, J.O. Hirschfelder, K.--H. Yang,
{\sl Rev. Mod. Phys.} {\bf 55}, 109 (1983).

\bibitem{Mermin} N.D. Mermin and H. Wagner, {\sl Phys. Rev. Lett.} {\bf 17}
(1966) 1133; S. Coleman, {\sl Commun. Math. Phys.} {\bf 31} (1973) 259.

\bibitem{Simon} B. Simon, {\sl Ann. Phys.} {\bf 97}, 279 (1976).

\bibitem{Perelomov} A.M.Perelomov and V.S.Popov, {\sl Theor. Math. Phys.}
{\bf 4}, 664 (1970).

\bibitem{Sal} P.Salabura {\em et al.}, {\sl Phys. Lett.} {\bf B245}, 153
(1990); I.Koenig {\em et al.}, {\sl Z. Phys.} {\bf A346}, 153 (1993).

\bibitem{Celenza} L.S.Celenza, V.K.Mishra, C.M.Shakin and K.F.Lin,
{\sl Phys. Rev. Lett.} {\bf 57}, 55 (1986); D.G.Caldi and A.Chodos,
{\sl Phys. Rev.} {\bf D36}, 2876 (1987); Y.J.Ng and Y.Kikuchi,
{\sl Phys. Rev.} {\bf D36}, 2880 (1987).

\bibitem{Caldi} D.G.Caldi and S.Vafaeisefat, {\cal Phys. Lett.} {\bf B287},
185 (1992).

\bibitem{Univ} T.Vachaspati, {\cal Phys. Lett.} {\bf B265}, 258 (1991).

\bibitem{Niel} N.K.Nielsen and P.Olesen, {\sl Nucl. Phys.} {\bf B144}, 376
(1978).

\bibitem{DESY} R.B\"{o}nish, DESY report 94-129, hep-th/9407162.

\end{thebibliography}
\end{document}